**Organic Carbon Cycle in the Atmosphere of Venus**
Jan Špaček

Firebird Biomolecular Sciences LLC,
13709 progress Boulevard, Alachua FL 32615
jspacek@firebirdbio.com

## ABSTRACT

The community commonly assumes that the Venusian atmosphere lacks organic (reduced) carbon. This is reflected in the literature, which now for almost a half a century does not mention organic carbon in connection with the Venus atmosphere. This assumption persists despite the failure of models that exclude organic carbon to account for many well-established observed features of the Venusian atmosphere.

Here is presented a model summarizing reactions that almost certainly are occurring in the Venusian atmosphere. The model relies on reactions of reduced carbon compounds known to occur in concentrated sulfuric acid (CSA), under high pressure and high temperature, many used in industrial processes. Inclusion of this known chemistry into a model for the Venus atmosphere accounts for several as yet unexplained observations. These include the upper haze, the lower haze, and the "mysterious" UV-blue absorber. This article also suggests, as a historical perspective, that a long-forgotten dispute is responsible for the persistent assumption of absence of interesting organic chemistry in the clouds of Venus.

Organic carbon certainly enters the Venus atmosphere via meteoric infall and possibly via other routes, discussed here. There, it encounters CSA aerosols in the Venusian clouds and undergoes evolution towards higher molecular weight, unsaturated, acid-soluble species. In the industrial literature, these are referred to as "red oil". The red oil evolves to absorb across larger parts of the UV and visible spectra, contributing to UV and visible darkening patterns in Venus at the cloud tops. As the sulfuric acid droplets containing red oil fall into hotter regions, sulfuric acid evaporates. Volatile alkenes are produced through cracking of the red oil and move up in altitude, where their condensate contributes to the upper haze. Since aromatic species in the red oil are resistant to cracking, residual organics yield highly porous aromatic carbon solid particles. These fall from the cloud and contribute to the lower haze. However, at the high temperature and pressure of the lower atmosphere, the aromatic carbon decomposes to give CO (the Boudouard reaction).

If the organic carbon comes only from meteoric infall and volcanism, the organic cycle would be insignificant in extent. However, if CO or $CO_2$ is reduced through photochemical or sulfuric acid-catalyzed processes, the extent of this organic carbon cycle would be larger, potentially explaining all of the unexplained phenomena mentioned above. Further, the cycle described by the model gives high molecular complexity, autocatalysis and selection for faster transformations. These are the elements of "life", although the logic of any biochemical metabolism based on it would be considerably different from the logic of biochemical metabolism operating in water in terran life.

### Highlights
Organics from meteorites enter Venus atmosphere.
Organic carbon polymerizes, forming unsaturated species in sulfuric acid.
Large unsaturated compounds contribute to observed UV absorption in Venus.
Two different organic classes contribute to upper and lower haze.

### 1. Introduction

Venus and its atmosphere have attracted renewed interest due to recent claims that the latter contains phosphine (Greaves et al., 2020) – a possible biosignature. This, in turn, revived the ongoing discussion of the possibility of life on Venus (Cockell, 1999; Limaye et al., 2018; Morowitz and Sagan, 1967; Seager et al., 2020). Life's origin, even in concentrated sulfuric acid (CSA, in this work meaning 70 – 98.3% in $H_2O$ by volume) forming clouds of Venus (Titov et al., 2018) presumably requires organic carbon chemistry in place.

Since the mid-1970s, the literature contains essentially no discussion of organics in the Venus atmosphere, even though they are certainly delivered by meteorites (Braukmüller et al., 2018). The lack of coverage of this topic in the literature is based on historical interpretations of observed data, which were affected by broader cultural circumstances (discussed in Section 4). As a result, the community appears to operate under an unwritten consensus that Venus lacks interesting organic chemistry (personal communication with the MIT Venus Cloud Life group; venuscloudlife.com).

Accordingly, astronomers try to interpret observations of Venus using inorganic species solely (Krasnopolsky, 2012, 2016; Titov et al., 2018). This despite the fact that current "inorganics only" models fail to account for many observations. These include the upper haze, the lower haze, and the "mysterious" UV-blue absorber.

Here, based on known reactivity of organic molecules in CSA, conversions of organic species introduced to Venus atmosphere are predicted. A partial list of reactions that organics undergo in CSA include: (a) protonation, (b) sulfonation, (c) hydrogen exchange, (d) dehydrogenation, (e) dehydration, (f) hydrogenation (if $H_2$ is present), (g) racemization, (h) isomerization, (i) creation of new C-C bonds to form larger molecules, and (j) C-C bond cleavage – (cracking; **Fig. 1**) (Bains et al., 2021a; Bains et al., 2021b; Baldwin, 1960; Burwell et al., 1954; Yong et al., 2008). Any hydrocarbons that might be present in the Venusian atmosphere would be exposed to CSA, and would undergo these reactions. The planet-wide bulk scale of these reactions might be limited if the only organic carbon were to be that delivered by meteorites from space (Braukmüller et al., 2018) and volcanism (Byrne et al., 2021; Filiberto et al., 2020; Marcq et al., 2013). However, they could be significant if carbon oxides were to be reduced and reintroduced into the organic carbon cycle (Section 2.5; **Fig. 3**).

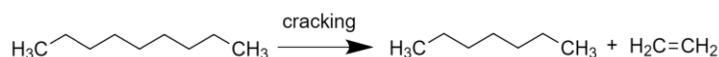

**Fig. 1.** *Schematic of hydrocarbon cracking*

Because CSA, high temperatures, and pressures are routinely used for chemical processing of organic compounds in industry (Brooks and Humphrey, 1918; Mianowski et al., 2012), a wealth of reliable literature describing these reactions can be used to model the behavior of reduced carbon in the Venus atmosphere. That this has not been done before in the analysis of Venus may reflect the gap separating those who practice astronomy and those who refine petroleum. The first understand (primarily) spectroscopy; the second have an intuitive understanding of reactivity of organics in CSA.

In developing a model for the Venusian atmosphere that includes organics, Section 1 describes reactions that are relevant for a hypothesized Venusian organic carbon cycle. Section 2 describes that carbon cycle and how species in the cycle contribute to yet unexplained phenomena observed in the Venus atmosphere. These include the identity of species forming upper and lower hazes and the species causing observed cloud coloration, UV absorption, and rapid changes in opacity. Section 3 briefly discusses the potential for life in the Venusian atmosphere based on the hypothesized cycle. Section 4 examines why the organic chemical cycles are to date rarely discussed for Venus.

The hypothesized model will likely have relevance to exoplanets as well as Venus. Recently, it was reported that exoplanets hosting liquid sulfuric acid may be as abundant as planets hosting liquid water (Ballesteros et al., 2019).

## 1.1. Sulfuric acid distillation

Pure CSA is a colorless liquid. However, traces of organic contaminants can turn it yellow. Darker, up to black, coloration appears upon storage of CSA contaminated with organics. Contaminated CSA is routinely distilled to increase its concentration and purity, as water and most of the impurities have lower boiling points than $H_2SO_4$ (Clark et al., 1990). During distillation, residual CSA turns from yellow to red-brown and black as the temperature increases. The changes in coloration depend more on the amount of the organics than on their chemical composition. Further heating turns the remaining CSA back to yellow or colorless, as the contaminants decompose to volatile species and are distilled off.

## 1.2. Carbon in concentrated sulfuric acid
### 1.2.1. Reactions in sulfuric acid

**Sugars**. To start with reactions familiar to students of chemistry, reactions with sugars involve predominantly dehydration and oxidation followed by polymerization and aromatization. During the oxidation $H_2SO_4$ and sugar react to produce $SO_2$, and 2 $H_2O$ and dehydrogenated species. These reactions turn the sugar-CSA mixture from colorless, to yellow, to brown, and to black. Both are exothermic and have faster kinetics at higher temperatures. With powdered sugar an explosive production of a "carbon snake" occurs as the heat produced during the reaction increases the reaction kinetics. A solid end product is black porous carbon, when the concentration of sugar in CSA is high. In solutions, where the final concentration of sugar in CSA is low, the reaction proceeds through yellow to rose-pink to brown coloration and finally to dark brown over a longer period of time. The darker colored compounds resulting from higher initial sugar concentrations were identified as "humic substances", while the yellow and red coloration is due to smaller conjugated molecules (Love, 1953).

**Hydrocarbons.** CSA polymerizes alkenes, converts them into secondary and tertiary alcohols, and yields esters of sulfuric acid (Brooks and Humphrey, 1918).

Paraffins (saturated hydrocarbons), normally unreactive in aqueous solutions, react in CSA (Eisner et al., 1940; Komarewsky and Ruther, 1950; Kramer, 1967). The reaction of saturated hydrocarbons is initiated by an oxidation during which $H_2SO_4$ is consumed, releasing $SO_2$ and 2 $H_2O$ molecules, resulting in alkenes. The originally acid-insoluble hydrocarbons are isomerized, polymerized, and converted into unsaturated colored acid soluble oils, which are in literature referred to as "conjugated polymers" (Miron and Lee, 1962) or "red oil" (Albright et al., 1972).

The formation of red oil is an undesired side reaction during industrial alkylation catalyzed by CSA; it can be mitigated, but not eliminated, by adjusting conditions (Huang et al., 2015). As a general rule, acidity below 91 % $H_2SO_4$ leads to oligomerization (Huang et al., 2015). At acidity below 85 % runaway autocatalytic side reactions proceed, where produced red oil catalyzes the conversion of paraffins to more red oil. If water content is kept below 1 % in CSA, up to 14 % of hydrocarbons can be accumulated in the CSA before the runaway occurs, rapidly increasing the red oil content (Albright et al., 1972). On the other hand, high acidity above 95 % of $H_2SO_4$ promotes cracking reactions (Huang et al., 2015). When stored, the red oil darkens at ambient temperature as its polymerization proceeds to larger, more conjugated molecules. The darkening is slowed at cold temperatures (Albright et al., 1972).

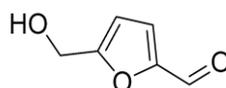

**Fig. 2**. 5-hydroxymethylfurfural, a dehydration product of sugar hexose.

**Particle formation.** When 5-hydroxymethylfurfural (**Fig. 2**) is added into heated CSA, black microporous particles are formed. These were identified as "humic particles", largely aromatic structures resulting from dehydration, dehydrogenation, carbon scaffold rearrangement, and sulfonation. With experimental conditions selected by Björnerbäck (2018), approximate chemical formula $C_{470}H_{213}O_{109}S$ was reported. These particles were found to have high surface area and high affinity towards $CO_2$ (Björnerbäck et al., 2018). Estes *et al.* (2019) showed that similar highly functionalized carbon nanoparticles can be formed by heating glycerol (Estes et al., 2019) and other alcohols and sugars in CSA (personal communication of corresponding author). The particle structure and chemical properties strongly depended on the temperature. The particles produced from glycerol had an O/C ratio up to 0.5. Its optical properties were similar to soot, graphene oxide, or humic acids, and included absorbance between 220-250 nm, extending to visible range, and weak fluorescence (Estes et al., 2019).

**"Red oil conversion."** The character of reactions of organic carbon in CSA strongly depends on the concentrations of sulfuric acid and the composition of the organic carbon reactants. Rates of various reactions are strongly influenced by temperature. Although the initial steps of the reactions depend on the structure of the initial reactants (collected by (Bains et al., 2021a; Bains et al., 2021b) in an excellent review), given enough time and organic carbon, the end products are visually similar. Reactions proceed through larger, more conjugated acid soluble polymers and towards aromatic black humic-like substances and solid carbon particles. In this work, this process of organic carbon polymerization in CSA from small organic molecules towards large polycyclic aromatic aggregate particles will be referred to as the "red oil conversion".

### 1.3. Optical properties of conjugated molecules

As organic carbon undergoes red oil conversion, forming progressively more conjugated larger molecules, its absorption spectrum changes. A conjugated system containing less than 8 double bonds generally absorbs only in the UV range and appears colorless. With every conjugated double bond added, the compound absorbs photons of longer wavelength, changing the compound's appearance from yellow to red and brown (Atkins and Paula, 2006). Finally, highly conjugated systems, such as humic acid-like substances (HULIS)(Graber and Rudich, 2006) or graphene-like substances, appear black, absorbing not only photons of visible parts of the spectrum, but also photons of broader UV and IR spectra (Andreae and Gelencsér, 2006; Dawlaty et al., 2008).

Highly conjugated molecules provide fluorescence (Thomas et al., 2007) including HULIS (Mobed et al., 1996), graphene oxide (Shang et al., 2012), and graphene (Zhao et al., 2017) with product specific excitation and emission profiles. Fluorescence of these substances can be used for their sensitive detection and approximate identification even when present in trace concentrations.

### 1.4. Production of red oil in Earth atmosphere

Interactions and reactions of organics and CSA aerosols are studied because of their relevance to Earth environment and climate. In the atmosphere CSA aerosols ("acid rain") are formed from $SO_2$ produced during fossil fuel burning.

An uptake of volatile organic compounds into CSA levitating droplets was observed (Lee et al., 2008) and it was shown that large multifunctional organic species are produced through acid catalysis from volatile organics (Jang et al., 2004).

In Earth's atmosphere, HULIS forms a large fraction of organic carbon present in the aerosols. Atmospheric HULIS is predominantly formed due to combustion of organic materials (Andreae and Gelencsér, 2006) but could be also formed as a secondary aerosol due to polymerization of isoprene and other alkenes in acidic droplets (Limbeck et al., 2003) producing UV absorbent fluorescent species (Hegglin et al., 2002). Although relevance of these observations to Earth atmosphere is debated (Lee et al., 2008) as humidity reduces efficiency of HULIS production in acidic droplets (Hegglin et al., 2002; Limbeck et al., 2003), these experimental data are relevant to conditions with lower humidity and higher CSA droplet concentration – such as can be found in the atmosphere of Venus.

## 1.5. High temperature carbon chemistry

This section is significant for understanding of the chemistry happening below the Venus cloud deck.

Both elevated temperature and high acidity promote cracking of aliphatic hydrocarbons (Brooks and Humphrey, 1918), while aromatic structures are resistant to cracking. At high temperatures sulfuric acid was shown to intercalate into graphite layers, separating the layers and making graphite more porous (Moissette et al., 1992)**.**

### 1.5.1.    Boudouard reaction

At high temperatures in $CO_2$ - CO mixed atmospheres, solid carbon and carbon oxides undergo the Boudouard reaction: $C_{solid} + CO_2 \leftrightarrow 2CO$ (Mianowski et al., 2015). The overall direction of the reaction is determined by the temperature, partial pressures of the oxide species, and total pressure. The gasification of solid carbon occurs in two steps: (1) $CO_2$ oxidizes the solid carbon, itself being reduced to CO. (2) The newly produced CO is released and leaves the solid carbon. The opposite process, formation of solid soot (coking) happens when the CO is above the equilibrium mixing value for given temperature and pressure. During the coking process, soot particles are formed in the atmosphere or on surfaces (Tangstad et al., 2018).

The Boudouard reaction products of carbon gasification have higher volume than the reactants. Therefore, one might expect the gasification to occur at higher temperatures at higher pressures. However, the opposite is true. The increased pressure increases the rate of the graphite oxidation and shifts the equilibrium to lower temperatures because the first step, oxidation of solid carbon by $CO_2$, is the rate determining step of the Boudouard reaction (Mianowski et al., 2012). If the oxidation step is facilitated by a catalyst or microwave radiation, the temperature of gasification is lowered by more than 400 °C (Hunt et al., 2013). Unsurprisingly, higher surface area of the solid carbon increases the overall reaction rate (Kaczorowski et al., 2007).

## 2.    The atmosphere of Venus

The Venusian atmosphere (**Fig. 3**) consists mostly of $CO_2$ and carries sedimenting particles from about 100 to about 10 km altitude (Titov et al., 2018). Its temperature and pressure range from +470 °C and 92 bars at the surface to approximately -100 °C and $2 \cdot 10^{-5}$ bar at the top of the upper haze (Seiff et al., 1985). At altitudes between ~48 to ~70 km (lower by about 4 km towards the higher latitudes) is a continuous cloud layer, mostly consisting of liquid sulfuric acid droplets with water as a minor component (Jessup et al., 2020; Oschlisniok et al., 2012; Ragent et al., 1985; Titov et al., 2018). The cloud appears to be to the naked eye "a pale lemon yellow" (Morowitz and Sagan, 1967; Ross, 1928) due to a yet unknown UV-blue absorber (Pollack et al., 1980). The cloud has large spatial and temporal variations in UV opacity (Lee et al., 2019). These variations are on a large scale dominated by Hadley and polar cells (Oschlisniok et al., 2021) and superrotation of the atmosphere (Lebonnois et al., 2010). The upper cloud is of photochemical origin, while the lower cloud forms due to $H_2SO_4$ and $H_2O$ condensation (Ragent et al., 1985). The identity of the particles forming the upper and lower haze remains unsolved (Jessup et al., 2020; Titov et al., 2018).

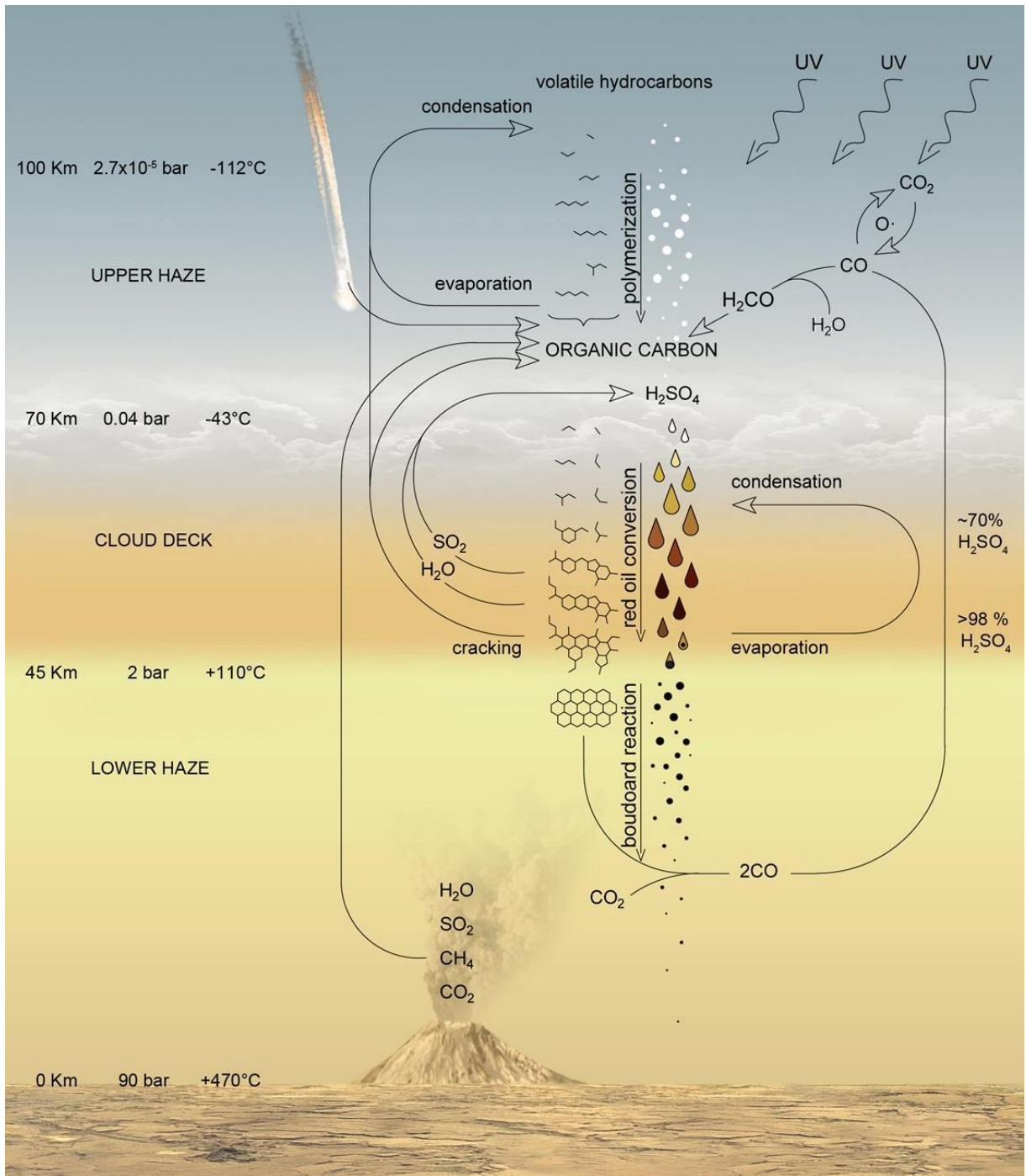

**Fig. 3.** Hypothetical carbon cycle in the equatorial region of the Venusian atmosphere. Organic carbon is introduced to the cycle via meteoric infall, volcanic activity, and processes (e.g. photochemistry) that deliver formaldehyde ($H_2CO$) or other organic carbon species. This organic carbon undergoes red oil conversion, producing larger conjugated and aromatic species, and $H_2O$ and $SO_2$ at the upper cloud undergo photochemical recombination to form $H_2SO_4$. Above 95 % sulfuric acid concentration (below ~50 km), aliphatic hydrocarbons undergo cracking and evaporate with other volatile molecules. Hydrocarbons condense forming aerosols at the upper atmosphere and undergo UV induced radical polymerization as they fall back to the cloud. There, they either contribute to the red oil production or undergo cracking at the lower atmosphere, returning to the upper haze. At the lowest part of the cloud water and subsequently sulfuric acid evaporate, increasing concentration of organic aromatic compounds which form porous carbon particles. Under the cloud carbon particles undergo Boudouard reaction and are converted to CO. (Photochemically) produced CO might be reduced at the upper atmosphere and reintroduced to the cycle or is oxidized by oxygen radical to $CO_2$. For simplicity winds aren't taken into account. Pressures and temperatures from (Seiff et al., 1985).

## 2.1. Cloud deck

The concentration of cloud deck sulfuric acid aerosol decreases with increasing altitude. At the bottom of the cloud deck, sulfuric acid is estimated to be above 100 % (a solution of $SO_3$ and $H_2S_2O_7$ in $H_2SO_4$). The $H_2SO_4$ concentration decreases roughly linearly from the cloud bottom to ~60 km elevation, where it reaches ~70 % $H_2SO_4$ in water. At higher altitudes, the $H_2SO_4$ is more concentrated (~75 %) in the upper cloud deck due to sharp drop in water vapor concentration and photochemical production of $H_2SO_4$ (Bains et al., 2021b; Titov et al., 2018). The temperature, linearly decreases from about 100 °C at the cloud bottom to -10 °C at 60 km and then drops more gradually to approximately -40 °C at 70 km altitude (Seiff et al., 1985).

These temperature profiles apply to equatorial regions (Seiff et al., 1985). With increasing latitude the cloud deck is shifted to lower altitude by about 4 km, following a decrease in the temperature (Titov et al., 2018). The 75 % CSA aerosols reportedly stay liquid down to -33 °C (Clapp et al., 1997). Liquid 75 % by weight (63 % by volume) sulfuric acid at -65 °C was reported by Tolbert et al. (1993). Hence, from the cloud bottom to ~65 km altitude (or potentially up to the cloud top) the aerosols are formed from liquid CSA, absorbing, dissolving, and concentrating organic compounds from the surrounding atmosphere (Tolbert et al., 1993). The organics dissolved in CSA droplets undergo reactions in a wide temperature, irradiation, and CSA concentration ranges as they are moved through the cloud.

The cloud deck is mixed by atmospheric superrotation and two major global scale meridional circulation systems on each hemisphere (Lebonnois et al., 2010). The larger Hadley cells rise near equator, flow poleward, downwell at mid latitudes and returns at low altitudes back towards the equator. The polar cell runs in the opposite direction, downwelling at mid latitude and upwelling at poles (Oschlisniok et al., 2021). The cloud is also mixed by localized eddies, downdrafts, gravitational waves, and upwellings (Baker et al., 1999). Finally, the aerosols are mixed by sedimentation. In general particles in the upper photochemically produced cloud region are long-lived, whereas the particles in the lower condensational layer may be recycled in hours (Ragent et al., 1985). Inevitably, all nonvolatile material falls out of the cloud deck over time, even though the particle lifetime can be extended by one of the above-described mixing processes.

## 2.2. Hypothesis: The unknown UV absorber is the red oil (organic carbon in CSA)

UV markings on the top of the clouds were for a long time the only distinctive feature of Venus (Knollenberg et al., 1977). Although the markings were known for nearly a century (Ross, 1928), the agent causing these is still unknown; literature still refers to an "unknown" or "mysterious" UV absorber present within cloud aerosol (Greaves et al., 2020; Pérez-Hoyos et al. 2018; Titov et al., 2018). The markings are detected at the cloud tops as a localized decrease in albedo ranging from 245 nm extending well past 500 nm into the blue portion of the spectrum (Jessup et al., 2020; Pérez-Hoyos et al., 2018). The markings are most pronounced close to the equator, forming large Y shapes, bands and streaks, variable on the order of hours to weeks. The biggest markings have the longest lifetime and highest UV contrast (Ragent et al., 1985). $SO_2$ contributes towards the UV absorption (band around 300 nm) and is often, but not always, collocated with the UV markings. Daytime Venus observations showed that $SO_2$ and the unknown UV absorber are both vertically transported to the cloud top through upwellings. While at the cloud tops the $SO_2$ undergoes fast photochemical destruction, while the UV absorber persists much longer (Jessup et al., 2020).

The localized upwellings appearing during daytime show as bright rimmed dark spots on the cloud tops. These patterns are created by a penetrative convective cell bringing up aerosols containing the UV absorber. The bright rims are hypothesized to be formed by the freshly photochemically produced $H_2SO_4$ aerosols caught in the downdraft, increasing the local albedo in the ring around the upwelled cell (Baker et al., 1999). The localized upwellings together with transport of the UV absorber from lower portions to the cloud tops via the Hadley cell, and gravitational waves, in combination with the super rotation of the atmosphere explain observed dark streaks localized at the cloud tops (Jessup et al., 2020). Observations of 365 nm UV opacity show that the unknown UV absorber is much less prevalent at polar regions (Molaverdikhani et al., 2012). The fact that significant amounts of the UV absorber are not brought up by polar upwellings indicates that the UV absorber is correlated more to photochemistry than to chemistry happening below the clouds.

The modeled rapid change of volumetric solar heating rate predicted by Baker (1999) causing the formation of the upwellings cells could be enhanced by following positive feedback mechanism: As small organic molecules dissolved to the cloud droplets fall to warmer regions, they increase their UV-blue absorbance due to the red oil conversion. After sunrise, droplets containing the UV absorber receive additional heating due to solar radiation, which in turn increases the red oil conversion rate, resulting in their faster darkening. This causes a positive feedback as darker droplets absorb more solar radiation and absorbed heat increases red oil conversion rate, making the droplets darker. Small initial variation in the region's opacity creates within hours a large difference in the solar heating rates of the atmospheric region. Further, as the dehydration and dehydrogenation are exothermic, chemical heating might contribute to overall heating (although this is likely a minor contribution to the overall heat balance and would strongly depend on the red oil content in the CSA aerosols). The heated regions become more buoyant and rise. The surrounding regions are pulled in to fill the displaced atmospheric mass, creating downdrafts surrounding the upwellings.

This hypothesis is supported by similarity of the absorbance spectrum profile of humic acid-like substances in CSA (Limbeck et al., 2003) and the broad absorption spectrum (245 nm – 500 nm) of the Venusian absorber (Jessup et al., 2020; Pérez-Hoyos et al., 2018). Note that the absorbance of the red oil in CSA increases and extends to the visible spectrum during its chemical evolution (Section 1.3). Hence if the unknown UV absorber is the red oil, shallower localized cloud upwellings will contain less polymerized red oil with UV absorbance predominantly in UV part of the spectrum, while absorbance spectrum of the UV absorber upwelled from lower parts of the cloud will extent well into the visible part of the spectrum.

### 2.3. Hypothesis: Organic carbon in the upper haze

Upper haze up to ~100 km altitude is present in all but polar regions of Venus. The upper haze is more absorbent in IR than in UV (Ragent et al., 1985).

Small hydrocarbons are formed during the red oil cracking. Due to their high volatility, these condense in the upper atmosphere, contributing to or forming the upper haze. The condensed hydrocarbon particles sediment through the atmosphere to the point where they sublimate or boil off.

Somewhat similar conditions are present on Titan, where hydrocarbons are known to undergo UV catalyzed polymerization towards large molecules often called tholins (Waite et al., 2007). Polymerization of light alkenes in the condensed particles is expected in the Venus atmosphere as well. The polymerized molecules, being less volatile, do not evaporate as readily and fall into the cloud deck. There, they are either dissolved in sulfuric acid droplets, re-entering the red oil conversion cycle, or sediment towards lower altitudes where they evaporate or undergo cracking.

Presence of (polymerized) hydrocarbons in the upper haze should be possible to detect via analysis of the Venus IR spectrum.

### 2.4. Hypothesis: Organic carbon in the lower haze

The lower haze starts at the bottom of the cloud deck, with some evidence for occasional haze down to 10 km altitude. It consists of a large quantity of small particles (r <0.25 μm), with size decreasing with decreasing altitude (Ragent et al., 1985).  Gnedykh et al. (1987; in (Titov et al., 2018)) suggested a lower haze mass density of 0.1-2 mg/m$^3$ at elevations above 35 km, with particle refractive index 2. This high refractive index was attributed to melted elemental sulfur in the particles (Sasson et al., 1985). Elemental sulfur droplets are supported by Krasnopolsky's model of Venus atmosphere (Krasnopolsky, 2016).

The model presented here attributes at least some of the lower haze to highly aromatic solid carbon particles (Björnerbäck et al., 2018; Estes et al., 2019), which must be falling from the clouds, as aromatic residues produced during red oil conversion are unable to undergo cracking and/or distillation under the conditions in the Venusian cloud. These highly porous carbon particles would sediment slower than liquid elemental sulfur droplets. The refractive index of soot depends on the H/C ratio (Habib and Vervisch, 1988); it is 1.96 for graphene oxide, and1.99 for thermally reduced graphene oxide, both of which match well the nephelometric observations of the lower haze, which has a refractive index 2 (Titov et al., 2018).

The observed global and temporal variation in lower haze particle density (Ragent et al., 1985) should be compared and matched to local production of the red oil in the cloud deck. The lower haze particles are likely to be weakly fluorescent, similarly to particles produced from glycerol in CSA (Estes et al., 2019).

The thermodynamics of the Boudouard reaction tells us that solid carbon cannot exist on the surface of Venus (Mianowski et al., 2015). At 470 °C and 90 bars in a predominantly $CO_2$ atmosphere (Seiff et al., 1985), all graphite and other forms of $sp^2$ carbon species are turned into CO via the reaction with $CO_2$ (Mianowski et al., 2015). Hence, porous carbon particles falling through the Venus atmosphere will be turned into CO gas at some point. However, evidence shows that the particles forming the lower haze decompose before 35 km altitude, at a temperature around 200 °C (Titov et al., 2018) – a temperature too low for gasification of graphite in the $CO_2$ atmosphere. As was shown in section 1.5.1., the rate limiting step in the Boudouard reaction is graphite oxidation. The CO release from oxidized graphite happens at much lower temperatures (Hunt et al., 2013). Further, the carbon particles formed in sulfuric acid are oxidized to a high degree with O/C ratio up to 0.5 (Björnerbäck et al., 2018; Estes et al., 2019). The Boudouard reaction could be further facilitated if the solid carbon were to be oxidized by a stronger oxidant than $CO_2$. Both sulfuric acid (Moissette et al., 1992) and $SO_3$ (Feicht and Breu, 2015) are known to intercalate into the graphite and the carbon particles produced in CSA have high affinity towards $CO_2$. Finally, formation of carbocations by microwaves was shown to considerably lower temperatures needed for carbon gasification (Hunt et al., 2013). Conceivably sulfuric acid intercalated in the porous carbon particles would not be fully evaporated below the clouds and could promote formation of the carbocations, lowering the temperature required for the first step of the Boudouard reaction. Therefore, it is conceivable that the majority of carbon particles in the lower haze are completely converted into the CO gas before falling below 35 km altitude.

## 2.5. Opened and closed carbon cycle

Organic carbon enters the Venus atmosphere through meteoric infall and, potentially, volcanic activity. It dissolves in sulfuric acid droplets in the Venusian clouds. The rate of red oil conversion is negligible at the cloud tops due to low temperature. The reaction rate increases with increasing temperature, resulting in a gradient of UV-blue opacity with decreasing altitude within the cloud deck. Part of the red oil produced undergoes cracking to light alkenes at the bottom of the cloud. These condense at high altitude, contributing to the upper haze. There, they undergo UV induced radical polymerization and fall back to the clouds, enter the CSA droplets and undergo further conversion. Part of the red oil produces larger polycyclic aromatic compounds, which upon CSA evaporation from the aerosols form porous particles falling from the clouds. Below the clouds, the carbon particles contribute to the lower haze and then are decomposed through Boudouard reaction to CO. The lifetime of the red oil in the Venus cloud is determined by lifetimes of the CSA aerosol droplets. Most of the red oil is turned to CO within months.

If the meteoric infal is the only source of organic carbon in the Venus atmosphere, the organic carbon cycle would be an insignificant part of Venusian atmosphere chemistry.

For this cycle to be a significant contributor to the Venusian atmospheric processes, carbon oxides must be recycled through their reduction – either within lithosphere (Section 2.5.1) or in atmosphere (Section 2.5.2), closing the cycle as depicted on Fig. 3.

The large amounts of the UV absorber throughout the cloud deck make the recycling scenario plausible. But it awaits confirmation by further experiment.

### 2.5.1.    Venusian volcanic activity

The concentration of $SO_2$ at the cloud tops spiked and then decreased by an order of magnitude during the 1970s and 1980s from ~500 ppm to ~50 ppm (Esposito, 1984). Similar spikes and decreases were observed again between 2007 and 2012. Both spikes in $SO_2$ concentrations were attributed to volcanic activity followed by long term $SO_2$ decomposition (Marcq et al., 2013). The hypothesis of active Venusian lithosphere was also proposed for repeated increase in the submicron haze density (Esposito, 1984). This hypothesis was supported by a recent proposal of active plate tectonics (Byrne et al., 2021), and observations of fresh unoxidized olivine in lava on the Venus surface (Filiberto et al., 2020). If Esposito (1984), Filiberto et al. (2020) and Byrne et al. (2021) are correct in their predictions and Venus is currently geologically active, the overall carbon and sulfur cycles on Venus potentially involve the lithosphere, supplying Venusian atmosphere with reduced carbon species.

### 2.5.2.    Carbon reduction in Venus atmosphere

Several pathways of carbon oxides reduction are tentatively proposed here.

Acid and transitional metals catalyzed hydrogenation of CO would be the most direct way for production of organic carbon (Rybacka et al., 2017) as CSA and likely transitional metals are present in the Venusian atmosphere (Krasnopolsky, 2017). However, lack of hydrogen gas in Venusian atmosphere indicates that amounts of thusly produced formaldehyde ($H_2CO$) would be insignificant.

Another option of $H_2CO$ production is through a series of photochemical reactions as suggested by Pinto et al. (1980) for early Earth. These rely on photochemical dissociation of $H_2O$ as a source of hydrogen atoms for reduction of CO. This process was proposed to produce significant amounts of $H_2CO$ on young Earth.

Recently a singlet pathway for formation of $O_2$ and C atoms from $CO_2$ has been proposed and shown to proceed with ~5 % yield after irradiation with ~105 nm UV light (Lu et al., 2014). The highly reactive carbon atoms produced in Venusian stratosphere by this process could subsequently combine with $H_2O$ to form $H_2CO$ or other reduced carbon species, avoiding a need for catalysts typically needed for photochemical reduction of $CO_2$ by $H_2O$ (Teramura and Tanaka, 2018).

It was shown that formaldehyde gas is rapidly and irreversibly dissolved by cold -40°C, CSA at 1 bar as well as at Earth stratospheric pressures (Tolbert et al., 1993). These conditions correlate well with conditions of Venus upper cloud, showing that $H_2CO$ potentially produced there is efficiently absorbed and concentrated in the forming CSA droplets.

### 3.    Potential for life in the Venus atmosphere with the carbon cycle

As noted by Bains *et al.* (2021b), life in the clouds of Venus cannot be based on nucleic acids, proteins, lipids, and sugars. All of these molecules, including their building blocks, are rapidly converted to red oils in CSA. The naturally occurring molecules reported as stable in clean CSA by themselves would react in more complex mixtures with carbocations generated in the red oil. As aliphatic hydrocarbons are not chemically stable above 70 % $H_2SO_4$, chemically stable lipid bilayers could not be formed. Hence, life in the Venusian cloud using different solvents than CSA separated from the outside environment with a membrane as suggested in Bains et al. (2020) is unlikely.

Red oil conversion gives highly complex carbon molecules, a small fraction of which can by chance acquire functionalities necessary for biochemistry (Bains et al., 2021b). Red oil is in principle autocatalytic (not self-replicating), causing runaway reactions producing exponentially more red oil in industrial processes (Albright et al., 1972). A form of the red oil with an above average rate of autocatalysis is conceivable.

All of the particles containing red oil eventually sediment below the cloud where the red oil is decomposed to CO at high temperature. But the red oils with higher rates of autocatalysis would more likely cause the upwelling cell in the cloud (see section 2.2), extending its lifetime in the cloud, potentially allowing longer survival of larger non-volatile molecules.

If self-replication of some carbon species is feasible within the Venus clouds, the progeny of these molecules would be more consistently elevated to the upper cloud layer due to solar heating, restarting the cycle and avoiding falling towards the lower haze by producing more of itself. The efficiently self-replicating species cause the upwelling cloud cells; neighboring less efficient red oils end up being pulled down in a downdraft, decomposing under the cloud to produce "nutrients". Selection in this competition would be for faster conversion to light absorbing species and for seeding as many droplets as possible.

If molecules capable of imprecise self-replication are competing for resources in the Venus cloud, they are "life" by some definitions (Trifonov, 2011). But without a deeper understanding of the complex red oil chemistry, we would not be able to distinguish them from the "regular red oil", which is not *self*-replicating. Even if some molecules in the Venus atmosphere are undergoing this type of Darwinian evolution, a consensus about their "life" status would be hard to reach, since this "life" would be too different from the one example we have here on Earth.

The overall high reactivity of all classes of molecules in a mixture with reactive carbocations in CSA might be an insurmountable obstacle for the life origin and continuation in Venus clouds. Further, here proposed hypothesis for "life" in the Venusian cloud lacks a mechanism for "seeding" of neighboring droplets with the self-reproducing species. If a mechanism for exchange of large molecules between the droplets does not exist, any form of life could not evolve in the Venus atmosphere.

Further study of organic carbon chemistry in CSA is needed as it has a broader exobiological relevance, given that planets with liquid sulfuric acid are predicted to be as common as planets with liquid water in our galaxy (Ballesteros et al., 2019).

## 4. Why the Venus carbon cycle is not prominently discussed today

Literature that describes reactions of organic carbon in CSA predates the discovery of Venus's cloud composition. From this literature, it is clear that the red oil conversion must be occurring in the Venus atmosphere as a chemical necessity, at least on a small scale due to meteoritic organic carbon infall. The question is: Why has the proposed organic carbon cycle not been discussed before, despite its strong support in industrial chemistry?

Organic molecules have indeed been proposed for Venus. In 1955, Fred Hoyle suggested that Venus is covered by an ocean of hydrocarbons (Hoyle, 1955; Pollack and Sagan, 1965). However, Hoyle was not the first to present the idea of hydrocarbons on Venus. Immanuel Velikovsky did so in his fiction "Worlds *in Collision"*, writing that Venus should be surrounded by a "blanket of hydrocarbons" (Velikovsky, 1950). His work was (correctly) criticized by the scientific community, but beloved and popular among the lay population.

This conflict culminated in 1974, when the American Association for the Advancement of Science arranged a meeting of Velikovsky and Carl Sagan. Sagan's criticism of Velikovsky's book is published (Goldsmith, 1979). Sagan and other scientists from the 1950s until the late 1970s dismissed everything that Velikovsky said in his book. Animosity towards Velikovsky can be sensed in the research from that time. For example, Plummer (1969) contradicted Velikovsky's conclusion about hydrocarbons in the Venus atmosphere by comparing reflectivity of several hydrocarbons with Earth-based observations of Venus infrared reflectivity. Because of the absence of the 2.4 µm wavelength dip in reflectivity, Plummer reported that hydrocarbons are not present in the Venus atmosphere, and concluded that the observed reflectivity could be fully explained by "very slender ice particles".

This incorrect conclusion aside, reexamination of the data presented in (Plummer, 1969) does not refute the presence of hydrocarbons in the Venus atmosphere. By higher modern standards, at best his data put an upper limit on hydrocarbon concentrations. In fact, some of the data included suggest the presence of the 2.4 μm wavelength dip in reflectivity, compatible with some hydrocarbon presence (Plummer, 1969).

In the 1960s, hydrocarbons were rejected as major constituents of the Venusian atmosphere by the scientific community based on theoretical calculations of thermodynamic equilibrium (Dayhoff et al., 1967, 1964). These estimates pointed out that the atmosphere is too oxidizing for the existence of hydrocarbons (Mueller, 1964). Analysis of radar reflectivity was also proposed to exclude substantial amounts of organics (Pollack and Sagan, 1965).

Since the late-1970s, hydrocarbons were concluded to not possibly be a *major* component of a Venus atmosphere and were not discussed in the literature since, despite never being rigorously refuted as a *minor* yet significant component. That is despite predictions of the stable existence of polycyclic aromatic compounds in the Venusian atmosphere (Dayhoff et al., 1964; Pollack and Sagan, 1965) and Florensky et al (1978) proposing that "brownish" metastable complex organic compounds cannot be ruled out.

Perhaps aided by Sagan's authority, the general consensus among astronomers was established in the 1980s that there is no interesting organic chemistry happening in the clouds of Venus. This consensus prevailed as a status quo among the Venus community until today.

## 5. Conclusion: summary of the model

Organic carbon is delivered constantly to the Venusian atmosphere through meteorite infall and potentially through volcanic activity. The organic carbon in Venus atmosphere undergoes red oil conversion (a series of reactions resulting in a mixture of large complex polyfunctional polyunsaturated highly aromatic molecules dissolved in CSA). Red oil conversion is one part of a larger organic carbon cycle happening in the Venusian atmosphere. Species generated in the carbon cycle contribute to, or are the main agents causing, the upper and lower haze, and the UV-blue spectrum darkening of the cloud. Hence the red oil potentially also explains the "pale lemon yellow" coloration of Venus poetically described by Carl Sagan. The red oil conversion could also contribute towards the $SO_2$ production and to the observed fast changing volumetric solar heating rates in the upwelling cells in the cloud deck. If a mechanism to reintroduce CO or $CO_2$ to the organic carbon cycle exists, as suggested in Section 2.5.1 and 2.5.2, the organic carbon, although still a minor component, might be a very significant part of the Venus atmosphere chemistry.

The red oil conversion produces large complex molecules with various functionalities, in theory capable of initiating and sustaining "life". However, the overall high reactivity of the organic compounds of the Venus cloud droplets and lack of known mechanism for spread of the "life" between the droplets makes life in Venus clouds unlikely.

Based on the points made in this work, the hypothesis of organic carbon as a minor yet significant component of the Venus atmosphere should be tested using modern methods. This model makes these testable predictions:

(1) The red oil is the unknown UV-blue absorber, it can be sensitively detected and identified via its fluorescence.
(2) The lower haze is formed by aromatic carbon particles produced via red oil conversion; these are weakly fluorescent, with slow sedimentation due to their high porosity.
(3) The lower haze is directly linked to the red oil production. Its density is correlated to the UV opacity of the region of the cloud from which the haze originated.
(4) Visible coloration of the aerosols at the bottom of the cloud deck depends on concentration of the red oil. It will range between yellow, brown, up to black in the equatorial to mid latitudes.
(5) The cloud top absorbance spectrum of markings caused by upwellings will depend on the depth of the aerosol origin. Shallower cloud upwellings will absorb only in the UV part of the spectrum, while upwellings from lower parts of the cloud deck will result in darkening extending to the visible part of the spectrum.

(6) The upper haze is at least partially caused by volatile hydrocarbons and their polymers, which should be detectable through their IR spectral signatures.

(7) The chemistry described here is experimentally reproducible under laboratory conditions.

If predictions (1) and (2) are tested using a falling probe measuring laser-induced fluorescence, special care must be made to prevent false positive signals. The heat shield heated during the atmospheric reentry might be a significant source of fluorescent compounds. An ablative heat shield impregnated with organic carbon (such as phenolic resin) should be avoided or at least ejected before the fluorescence measurements begin.

### Acknowledgements

I am indebted to the MIT / Breakthrough Initiative and Firebird Biomolecular Sciences LLC for providing support for this project. I am also indebted to many comments and suggestions by Steven Benner, Gage Owens, and the MIT Venus Cloud Life group, including Janusz Petkowsky, William Bains, Paul Rimmer, David Grinspoon, and Sara Seager.

### References

Albright, L.F., Houle, L., Sumutka, A.M., Eckert, R.E., 1972. Alkylation of Isobutane with Butenes: Effect of Sulfuric Acid Compositions. Ind. Eng. Chem. Process Des. Dev. 11, 446–450. https://doi.org/10.1021/i260043a020

Andreae, M.O., Gelencsér, A., 2006. Black carbon or brown carbon? the nature of light-absorbing carbonaceous aerosols. Atmos. Chem. Phys. 6, 3131–3148. https://doi.org/10.5194/acp-6-3131-2006

Atkins, P.W., Paula, J., 2006. Physical Chemistry - 8th Edition. W.H. Freeman & Company.

Bains, W., Petkowski, J.J., Seager, S., Ranjan, S., Sousa-Silva, C., Rimmer, P.B., Zhan, Z., Greaves, J.S., Richards, A.M.S., 2020. Phosphine on Venus Cannot be Explained by Conventional Processes. arXiv Prepr. 06499.

Bains, W., Petkowski, J.J., Seager, S. 2021a. A Data Resource for Sulfuric Acid Reactivity of Organic Chemicals. Data 6, 24. https://doi.org/10.3390/data6030024

Bains, W., Petkowski, J.J., Zhan, Z., Seager, S., 2021b. Evaluating alternatives to water as solvents for life: The example of sulfuric acid. Life 11. https://doi.org/10.3390/life11050400

Baker, R.D., Schubert, G., Jones, P.W., 1999. High Rayleigh number compressible convection in Venus' atmosphere: Penetration, entrainment, and turbulence. J. Geophys. Res. E Planets 104, 3815–3832. https://doi.org/10.1029/1998JE900029

Baldwin, W.H., 1960. The Reaction Between Sulfuric Acid and Aliphatic Hydrocarbons. AMSCO 125- 82 Rep. https://doi.org/10.2172/4178737

Balla, A., Marcu, C., Axente, D., Borodi, G., Lazăr, D., 2012. Catalytic reduction of sulfuric acid to sulfur dioxide. Cent. Eur. J. Chem. 10, 1817–1823. https://doi.org/10.2478/s11532-012-0099-x

Ballesteros, F.J., Fernandez-Soto, A., Martínez, V.J., 2019. Diving into Exoplanets: Are Water Seas the Most Common? Astrobiology 19, 642–654. https://doi.org/10.1089/ast.2017.1720

Björnerbäck, F., Bernin, D., Hedin, N., 2018. Microporous Humins Synthesized in Concentrated Sulfuric Acid Using 5-Hydroxymethyl Furfural. ACS Omega 3, 8537–8545. https://doi.org/10.1021/acsomega.8b01274

Braukmüller, N., Wombacher, F., Hezel, D.C., Escoube, R., Münker, C., 2018. The chemical composition of carbonaceous chondrites: Implications for volatile element depletion, complementarity and alteration. Geochim. Cosmochim. Acta 239, 17–48. https://doi.org/10.1016/j.gca.2018.07.023

Brooks, B.T., Humphrey, I., 1918. The action of concentrated sulfuric acid on olefins, with particular reference to the refining of petroleum distillates. J. Am. Chem. Soc. 40, 822–856. https://doi.org/10.1021/ja02238a012

Burwell, R.L., Scott, R.B., Maury, L.G., Hussey, A.S., 1954. The Action of 96% Sulfuric Acid on Alkanes at 60°. J. Am. Chem. Soc. 76, 5822–5827. https://doi.org/10.1021/ja01651a085

Byrne, P.K., Ghail, R.C., Şengör, A.M.C., James, P.B., Klimczak, C., Solomon, S.C., 2021. A globally fragmented and mobile lithosphere on Venus. Proc. Natl. Acad. Sci. 118, e2025919118. https://doi.org/10.1073/pnas.2025919118

Clapp, M.L., Niedziela, R.F., Richwine, L.J., Dransfield, T., Miller, R.E., Worsnop, D.R., 1997. Infrared spectroscopy of sulfuric acid/water aerosols: Freezing characteristics. J. Geophys. Res. Atmos. 102, 8899–8907. https://doi.org/10.1029/97jd00012

Clark, R.S., Davison, J.B., Persichini, D.W., Yuan, W.I., Lipisko, B.A., Jones, A.W., Jones, A.H.J., Hoffman, J.G., 1990. Method for making ultrapure sulfuric acid.

Cockell, C.S., 1999. Life on Venus. Planet. Space Sci. 47, 1487–1501. https://doi.org/10.1016/S0032-0633(99)00036-7

Corgnale, C., Gorensek, M.B., Summers, W.A., 2020. Review of sulfuric acid decomposition processes for sulfur-based thermochemical hydrogen production cycles. Processes 8, 1–22. https://doi.org/10.3390/pr8111383

Dawlaty, J.M., Shivaraman, S., Strait, J., George, P., Chandrashekhar, M., Rana, F., Spencer, M.G., Veksler, D., Chen, Y., 2008. Measurement of the optical absorption spectra of epitaxial graphene from terahertz to visible. Appl. Phys. Lett. 93, 131905. https://doi.org/10.1063/1.2990753

Dayhoff, M.O., Eck, R. V., Lippincott, E.R., Sagan, C., 1967. Venus: Atmospheric evolution. Science. 155, 556–558. https://doi.org/10.1126/science.155.3762.556

Dayhoff, M.O., Lippincott, E.R., Eck, R. V., 1964. Thermodynamic equilibria in prebiological atmospheres. Science. 146, 1461–1464. https://doi.org/10.1126/science.146.3650.1461

Eisner, A., Fein, M.L., Fisher, C.H., 1940. Neutral Oils from Coal Hydrogenation. Action of sulfuric acid. Ind. Eng. Chem. 32, 1614–1621. https://doi.org/10.1021/ie50372a022

Esposito, L.W., 1984. Sulfur dioxide: Episodic injection shows evidence for active Venus volcanism. Science. 223, 1072–1074. https://doi.org/10.1126/science.223.4640.1072

Estes, C.S., Gerard, A.Y., Godward, J.D., Hayes, S.B., Liles, S.H., Shelton, J.L., Stewart, T.S., Webster, R.I., Webster, H.F., 2019. Preparation of highly functionalized carbon nanoparticles using a one-step acid dehydration of glycerol. Carbon N. Y. 142, 547–557. https://doi.org/10.1016/j.carbon.2018.10.074

Feicht, P., Breu, J., 2015. Gas-phase Preparation of SO3-Graphite: Host-Exchange and Exfoliation.

Zeitschrift fur Anorg. und Allg. Chemie 641, 1093–1098. https://doi.org/10.1002/zaac.201500168

Filiberto, J., Trang, D., Treiman, A.H., Gilmore, M.S., 2020. Present-day volcanism on Venus as evidenced from weathering rates of olivine. Sci. Adv. 6, 7445. https://doi.org/10.1126/sciadv.aax7445

Florensky, C. P., Volkov, V. P., Nikolaeva, O. V. 1978. A geochemical model of the Venus troposphere. Icarus, 33, 537-553 https://doi.org/10.1016/0019-1035(78)90189-6

Gao, P., Zhang, X., Crisp, D., Bardeen, C.G., Yung, Y.L., 2014. Bimodal distribution of sulfuric acid aerosols in the upper haze of Venus. Icarus 231, 83–98. https://doi.org/10.1016/j.icarus.2013.10.013

Goldsmith, D., 1979. Scientists Confront Velikovsky. Norton.

Graber, E.R., Rudich, Y., 2006. Atmospheric HULIS: How humic-like are they? A comprehensive and critical review. Atmos. Chem. Phys. 6, 729–753. https://doi.org/10.5194/acp-6-729-2006

Greaves, J.S., Richards, A.M.S., Bains, W., Rimmer, P.B., Sagawa, H., Clements, D.L., Seager, S., Petkowski, J.J., Sousa-Silva, C., Ranjan, S., Drabek-Maunder, E., Fraser, H.J., Cartwright, A., Mueller-Wodarg, I., Zhan, Z., Friberg, P., Coulson, I., Lee, E., Hoge, J., 2020. Phosphine gas in the cloud decks of Venus. Nat. Astron. 1–10. https://doi.org/10.1038/s41550-020-1174-4

Habib, Z.G., Vervisch, P., 1988. On The Refractive Index of Soot at Flame Temperature. Combust. Sci. Technol. 59, 261–274. https://doi.org/10.1080/00102208808947100

Hegglin, M.I., Krieger, U.K., Koop, T., Peter, T., 2002. Technical note: Organics-induced fluorescence in Raman studies of sulfuric acid aerosols. Aerosol Sci. Technol. 36, 510–512. https://doi.org/10.1080/027868202753571331

Hoyle, F., 1955. Frontiers of astronomy a: revolutionary new view of the universe. Harper & Brothers, New York.

Huang, Q., Zhao, G., Zhang, S., Yang, F., 2015. Improved catalytic lifetime of H2SO4 for isobutane alkylation with trace amount of ionic liquids buffer. Ind. Eng. Chem. Res. 54, 1464–1469. https://doi.org/10.1021/ie504163h

Hunt, J., Ferrari, A., Lita, A., Crosswhite, M., Ashley, B., Stiegman, A.E., 2013. Microwave-specific enhancement of the carbon-carbon dioxide (boudouard) reaction. J. Phys. Chem. C 117, 26871–26880. https://doi.org/10.1021/jp4076965

Imamura, T., Hashimoto, G.L., 2001. Microphysics of Venusian clouds in rising tropical air. J. Atmos. Sci. 58, 3597–3612. https://doi.org/10.1175/1520-0469(2001)058<3597:MOVCIR>2.0.CO;2

Jang, M., Czoschke, N.M., Northcross, A.L., 2004. Atmospheric organic aerosol production by heterogeneous acid-catalyzed reactions. ChemPhysChem. https://doi.org/10.1002/cphc.200301077

Jessup, K.L., Marcq, E., Bertaux, J.L., Mills, F.P., Limaye, S., Roman, A., 2020. On Venus' cloud top chemistry, convective activity and topography: A perspective from HST. Icarus 335, 113372. https://doi.org/10.1016/j.icarus.2019.07.006

Kaczorowski, J., Lindstad, T., Syvertsen, M., 2007. The influence of potassium on the boudouard reaction in manganese production. ISIJ Int. 47, 1599–1604. https://doi.org/10.2355/isijinternational.47.1599

Knollenberg, R.G., Hansen, J., Ragent, B., Martonchik, J., Tomasko, M., 1977. The clouds of Venus. Space Sci. Rev. 20, 329–354. https://doi.org/10.1007/BF02186469

Komarowsky, V.I., Ruther, W.E., 1950. Reactions of Paraffin Hydrocarbons in the Presence of Sulfuric Acid. J. Am. Chem. Soc. 72, 5501–5503. https://doi.org/10.1021/ja01168a034

Kramer, G.M., 1967. The oxidation of paraffins in Sulfuric Acid. J. Org. Chem. 32, 1916–1918. https://doi.org/10.1016/S0167-2991(08)64840-5

Krasnopolsky, V.A., 2016. Sulfur aerosol in the clouds of Venus. Icarus 274, 33–36. https://doi.org/10.1016/j.icarus.2016.03.010

Krasnopolsky, V.A., 2015. Vertical profiles of H2O, H2SO4, and sulfuric acid concentration at 45-75 km on Venus. Icarus 252, 327–333. https://doi.org/10.1016/j.icarus.2015.01.024

Krasnopolsky, V.A., 2013. S3 and S4 abundances and improved chemical kinetic model for the lower atmosphere of Venus. Icarus 225, 570–580. https://doi.org/10.1016/j.icarus.2013.04.026

Krasnopolsky, V.A., 2012. A photochemical model for the Venus atmosphere at 47-112km. Icarus 218, 230–246. https://doi.org/10.1016/j.icarus.2011.11.012

Krasnopolsky, V.A., 2007. Chemical kinetic model for the lower atmosphere of Venus. Icarus 191, 25–37. https://doi.org/10.1016/j.icarus.2007.04.028

Lebonnois, S., Hourdin, F., Eymet, V., Crespin, A., Fournier, R., Forget, F., 2010. Superrotation of Venus' atmosphere analyzed with a full general circulation model. J. Geophys. Res. E Planets 115. https://doi.org/10.1029/2009JE003458

Lee, A.K.Y., Li, Y.J., Lau, A.P.S., Chan, C.K., 2008. A Re-Evaluation on the Atmospheric Significance of Octanal Vapor Uptake by Acidic Particles: Roles of Particle Acidity and Gas-Phase Octanal Concentration. https://doi.org/10.1080/02786820802382736 42, 992–1000. https://doi.org/10.1080/02786820802382736

Lee, Y.J., Jessup, K.-L., Perez-Hoyos, S., Titov, D. V., Lebonnois, S., Peralta, J., Horinouchi, T., Imamura, T., Limaye, S., Marcq, E., Takagi, M., Yamazaki, A., Yamada, M., Watanabe, S., Murakami, S., Ogohara, K., McClintock, W.M., Holsclaw, G., Roman, A., 2019. Long-term Variations of Venus's 365 nm Albedo Observed by Venus Express , Akatsuki , MESSENGER , and the Hubble Space Telescope. Astron. J. 158, 126. https://doi.org/10.3847/1538-3881/ab3120

Lewis, J.S., 1974. Volatile element influx on Venus from cometary impacts. Earth Planet. Sci. Lett. 22, 239–244. https://doi.org/10.1016/0012-821X(74)90087-9

Limaye, S.S., Mogul, R., Smith, D.J., Ansari, A.H., Słowik, G.P., Vaishampayan, P., 2018. Venus' spectral signatures and the potential for life in the clouds. Astrobiology 18, 1181–1198. https://doi.org/10.1089/ast.2017.1783

Limbeck, A., Kulmala, M., Puxbaum, H., 2003. Secondary organic aerosol formation in the atmosphere via heterogeneous reaction of gaseous isoprene on acidic particles. Geophys. Res. Lett. 30. https://doi.org/10.1029/2003GL017738

Love, R.M., 1953. Spectroscopic studies of carbohydrates. 1. The action of sulphuric acid on sugars. Biochem. J. 55, 126–132. https://doi.org/10.1042/bj0550126

Lu, Z., Chang, Y.C., Yin, Q.-Z., Ng, C.Y., Jackson, W.M., 2014. Evidence for direct molecular oxygen production in CO2 photodissociation. Science. 346, 61–64. https://doi.org/10.1126/SCIENCE.1257156

Marcq, E., Bertaux, J.L., Montmessin, F., Belyaev, D., 2013. Variations of sulphur dioxide at the cloud top of Venus's dynamic atmosphere. Nat. Geosci. 6, 25–28. https://doi.org/10.1038/ngeo1650

Marcq, E., Lea Jessup, K., Baggio, L., Encrenaz, T., Lee, Y.J., Montmessin, F., Belyaev, D., Korablev, O., Bertaux, J.L., 2020. Climatology of SO2 and UV absorber at Venus' cloud top from SPICAV-UV nadir dataset. Icarus 335, 113368. https://doi.org/10.1016/j.icarus.2019.07.002

Marcq, E., Mills, F.P., Parkinson, C.D., Vandaele, A.C., 2018. Composition and Chemistry of the Neutral Atmosphere of Venus, Space Science Reviews. https://doi.org/10.1007/s11214-017-0438-5

Mianowski, A., Radko, T., Siudyga, T., 2015. The reactivity of cokes in Boudouard–Bell reactions in the context of an Ergun model. J. Therm. Anal. Calorim. 2015 1222 122, 1013–1021. https://doi.org/10.1007/S10973-015-4761-3

Mianowski, A., Robak, Z., Tomaszewicz, M., Stelmach, S., 2012. The Boudouard-Bell reaction analysis under high pressure conditions, in: Journal of Thermal Analysis and Calorimetry. Springer, pp. 93–102. https://doi.org/10.1007/s10973-012-2334-2

Miron, S., Lee, R.J., 1962. Molecular structure of conjunct polymers. Ind. Eng. Chem. Prod. Res. Dev. 1, 287–289. https://doi.org/10.1021/i360004a016

Mobed, J.J., Hemmingsen, S.L., Autry, J.L., Mcgown, L.B., 1996. Fluorescence characterization of IHSS humic substances: Total luminescence spectra with absorbance correction. Environ. Sci. Technol. 30, 3061–3065. https://doi.org/10.1021/es960132I

Moissette, A., Fuzellier, H., Burneau, A., Dubessy, J., Guérard, D., Lelaurain, M., 1992. Spontaneous Intercalation of Sulfuric Acid into Graphite and Synthesis of a New Graphite-Uranyl Sulfate Compound. Mater. Sci. Forum 91–93, 95–99. https://doi.org/10.4028/www.scientific.net/msf.91-93.95

Molaverdikhani, K., McGouldrick, K., Esposito, L.W., 2012. The abundance and vertical distribution of the unknown ultraviolet absorber in the venusian atmosphere from analysis of Venus Monitoring Camera images. Icarus 217, 648–660. https://doi.org/10.1016/j.icarus.2011.08.008

Morowitz, H., Sagan, C., 1967. Life in the clouds of venus? Nature. https://doi.org/10.1038/2151259a0

Mueller, R.F., 1964. A chemical model for the lower atmosphere of Venus. Icarus 3, 285–298. https://doi.org/10.1016/0019-1035(64)90037-5

Oschlisniok, J., Häusler, B., Pätzold, M., Tellmann, S., Bird, M.K., Peter, K., Andert, T.P., 2021. Sulfuric acid vapor and sulfur dioxide in the atmosphere of Venus as observed by the Venus Express radio science experiment VeRa. Icarus 362, 114405. https://doi.org/10.1016/j.icarus.2021.114405

Oschlisniok, J., Häusler, B., Pätzold, M., Tyler, G.L., Bird, M.K., Tellmann, S., Remus, S., Andert, T., 2012. Microwave absorptivity by sulfuric acid in the Venus atmosphere: First results from the Venus Express Radio Science experiment VeRa. Icarus 221, 940–948. https://doi.org/10.1016/j.icarus.2012.09.029

Pérez-Hoyos, S., Sánchez-Lavega, A., García-Muñoz, A., Irwin, P.G.J., Peralta, J., Holsclaw, G., McClintock, W.M., Sanz-Requena, J.F., 2018. Venus Upper Clouds and the UV Absorber From MESSENGER/MASCS Observations. J. Geophys. Res. Planets 123, 145–162. https://doi.org/10.1002/2017JE005406

Pinto, J.P., Gladstone, G.R., Yung, Y.L., 1980. Photochemical production of formaldehyde in Earth's primitive atmosphere. Science. 210, 183–185. https://doi.org/10.1126/science.210.4466.183

Plummer, W.T., 1969. Venus clouds: Test for hydrocarbons. Science. 163, 1191–1192. https://doi.org/10.1126/science.163.3872.1191

Pollack, J.B., Sagan, C., 1965. The microwave phase effect of Venus. Icarus 4, 62–103. https://doi.org/10.1016/0019-1035(65)90018-7

Pollack, J.B., Toon, O.B., Whitten, R.C., Boese, R., Ragent, B., Tomasko, M., Esposito, L., Travis, L., Wiedman, D., 1980. Distribution and source of the UV absorption in Venus' atmosphere. J. Geophys. Res. 85, 8141. https://doi.org/10.1029/ja085ia13p08141

Ragent, B., Esposito, L.W., Tomasko, M.G., Marov, M.Y., Shari, V.P., Lebedev, V.N., 1985. Particulate matter in the Venus atmosphere. Adv. Sp. Res. 5, 85–115. https://doi.org/10.1016/0273-1177(85)90199-1

Ross, F. E., 1928. Photographs of Venus, Astrophys. J. 68, 57. https://doil.org/10.1086/143130

Rybacka, O., Czapla, M., Skurski, P., 2017. Mechanisms of carbon monoxide hydrogenation yielding formaldehyde catalyzed by the representative strong mineral acid, H2SO4, and Lewis-Brønsted superacid, HF/AlF3. Phys. Chem. Chem. Phys. 19, 18047–18054. https://doi.org/10.1039/c7cp03362a

Sasson, R., Wright, R., Arakawa, E.T., Khare, B.N., Sagan, C., 1985. Optical properties of solid and liquid sulfur at visible and infrared wavelengths. Icarus. https://doi.org/10.1016/0019-1035(85)90061-2

Seager, S., Petkowski, J.J., Gao, P., Bains, W., Bryan, N.C., Ranjan, S., Greaves, J., 2020. The Venusian Lower Atmosphere Haze as a Depot for Desiccated Microbial Life: A Proposed Life Cycle for Persistence of the Venusian Aerial Biosphere. Astrobiology 21, 1–18. https://doi.org/10.1089/ast.2020.2244

Seiff, A., Schofield, J.T., Kliore, A.J., Taylor, F.W., Limaye, S.S., Revercomb, H.E., Sromovsky, L.A., Kerzhanovich, V. V., Moroz, V.I., Marov, M.Y., 1985. Models of the structure of the atmosphere of Venus from the surface to 100 kilometers altitude. Adv. Sp. Res. 5, 3–58. https://doi.org/10.1016/0273-1177(85)90197-8

Shang, J., Ma, L., Li, J., Ai, W., Yu, T., Gurzadyan, G.G., 2012. The origin of fluorescence from graphene oxide. Sci. Rep. 2. https://doi.org/10.1038/srep00792

Svedhem, H., Titov, D., Taylor, F., Witasse, O., 2009. Venus express mission. J. Geophys. Res. E Planets 114, 0–33. https://doi.org/10.1029/2008JE003290

Tangstad, M., Beukes, J.P., Steenkamp, J., Ringdalen, E., 2018. Coal-based reducing agents in ferroalloys and silicon production, in: New Trends in Coal Conversion: Combustion, Gasification, Emissions, and Coking. Elsevier, pp. 405–438. https://doi.org/10.1016/B978-0-08-102201-6.00014-5

Teramura, K., Tanaka, T., 2018. Necessary and sufficient conditions for the successful three-phase photocatalytic reduction of CO2 by H2O over heterogeneous photocatalysts. Phys. Chem. Chem. Phys. 20, 8423–8431. https://doi.org/10.1039/C7CP07783A

Thomas, S.W., Joly, G.D., Swager, T.M., 2007. Chemical sensors based on amplifying fluorescent conjugated polymers. Chem. Rev. https://doi.org/10.1021/cr0501339

Titov, D. V., Ignatiev, N.I., McGouldrick, K., Wilquet, V., Wilson, C.F., 2018. Clouds and Hazes of Venus.

Space Sci. Rev. https://doi.org/10.1007/s11214-018-0552-z

Tolbert, M.A., Pfaff, J., Jayaweera, I., Prather, M.J., 1993. Uptake of formaldehyde by sulfuric acid solutions: Impact on stratospheric ozone. J. Geophys. Res. Atmos. 98, 2957–2962. https://doi.org/10.1029/92JD02386

Trifonov, E.N., 2011. Vocabulary of definitions of life suggests a definition. J. Biomol. Struct. Dyn. 29, 259–266. https://doi.org/10.1080/073911011010524992

Truong, N., Lunine, J.I., 2021. Volcanically extruded phosphides as an abiotic source of Venusian phosphine. Proc. Natl. Acad. Sci. U. S. A. 118, e2021689118. https://doi.org/10.1073/pnas.2021689118

Velikovsky, I., 1950. Worlds in collision. Paradigma Ltd.

Waite, J.H., Young, D.T., Cravens, T.E., Coates, A.J., Crary, F.J., Magee, B., Westlake, J., 2007. Planetary science: The process of tholin formation in Titan's upper atmosphere. Science. 316, 870–875. https://doi.org/10.1126/science.1139727

Yong, J.L., Lee, A.K.Y., Lau, A.P.S., Chan, C.K., 2008. Accretion reactions of octanal catalyzed by sulfuric acid: Product identification, reaction pathways, and atmospheric implications. Environ. Sci. Technol. 42, 7138–7145. https://doi.org/10.1021/es7031373

Yung, Y.L., Demore, W.B., 1982. Photochemistry of the stratosphere of Venus: Implications for atmospheric evolution. Icarus 51, 199–247. https://doi.org/10.1016/0019-1035(82)90080-X

Zhao, S., Rondin, L., Delport, G., Voisin, C., Beser, U., Hu, Y., Feng, X., Müllen, K., Narita, A., Campidelli, S., Lauret, J.S., 2017. Fluorescence from graphene nanoribbons of well-defined structure. Carbon N. Y. 119, 235–240. https://doi.org/10.1016/j.carbon.2017.04.043